\documentclass[namedreferences]{SolarPhysics}
\usepackage{spr-sola-addons} 
\usepackage{epsfig}
\usepackage{lscape}          
\usepackage{url}             

\newcommand{\ion}[2]{#1\,{\small #2}}
\newcommand{\todash}{\,--\,}
\newcommand{\etal}{{\it et al.}}
\newcommand{\be}{\begin{equation}}
\newcommand{\ee}{\end{equation}}
\newcommand{\beq}{\begin{eqnarray}}
\newcommand{\eeq}{\end{eqnarray}}

\newcommand{\adv}[3]{   #1, {\it Adv. Space Res.}    {\bf  #2}, #3.}

\newcommand{\aspj}[3]{  #1, {\it Astrophys. J.} {\bf  #2}, #3.}

\newcommand{\sph}[3]{   #1, {\it Solar Phys.} {\bf  #2}, #3.}

\begin{document}
\begin{article}
\begin{opening}

\title{Solar Oscillation Frequency Changes on \\Time Scales of Nine Days} 

\author{S.\,C.~\surname{Tripathy}$^1$ \sep 
        F.~\surname{Hill}$^1$ \sep 
        K.~\surname{Jain}$^1$ \sep
        J.\,W.~\surname{Leibacher}$^{1,2,3}$
        }
\runningauthor{S. C. Tripathy \etal }
\runningtitle{Oscillation Frequency Changes on Short Time Scales}

\institute{$^{1}$GONG Program, National Solar Observatory, 950 N. Cherry Avenue, Tucson, 
AZ 85719, USA \\ {email: stripathy@nso.edu}\\
$^{2}$ Institut d'Astrophysique Spatiale (CNRS), Universit\'e
Paris-Sud~11, Orsay, France\\ 
$^{3}$ Observatoire de Paris, LESIA (CNRS),
                  F-92195 Meudon Principal Cedex, France}
\date{Received: date / Accepted: date}

\begin{abstract}
We establish that global solar $p$-mode frequencies can be measured
with sufficient  precision on time scales as short as
nine days to detect activity-related shifts. Using ten years of GONG data, we report that  
mode-mass and error-weighted frequency shifts derived from nine days
are significantly correlated with the strength of solar activity and
are consistent with long-duration measurements from GONG and the SOHO/MDI
instrument. The analysis of the year-wise distribution of the frequency shifts with change in  activity 
indices shows that both the 
linear-regression slopes and the magnitude of the  
correlation varies from year to year and they 
are well correlated with each other.  The study also indicates that the
magnetic indices behave differently in the rising and falling phases of
the activity cycle. For the short-duration nine-day observations, we
report  a higher sensitivity to activity.

\end{abstract}
\keywords{Sun: activity, Sun: oscillations,  Sun: helioseismology}
\end{opening}

\section{Introduction}
\label{Introduction} 
It is now well established that the global oscillation frequencies of
the Sun change in phase with the solar activity cycle, see for example
Jain and Bhatnagar (2003)  for intermediate-degree modes,
Jim{\'e}nez-Reyes \emph{et al.} (2004) and Howe \emph{et al.} (2006)
for low-degree modes, and references therein.  However, there is still
no consensus regarding the physical mechanism that gives rise to these
changes (Kuhn, 2001).  One deficiency is the lack of measurements for
modes at high frequency and high degree (Rhodes, Reiter, and Schou,
2003) and little information about the correlation on short time
scales.  Since the {\it p}-mode frequency changes are thought to be
associated with individual active regions that come and go continuously
(Hindman \etal, 2000), one would anticipate that the frequencies also
change continuously on any time scale. Due to the  finite lifetime of
the modes, the correlation between frequency and activity may depend on
the length of the observing run.  Thus, the use of mode frequencies and
other parameters derived from short time series, during which the solar
activity varies less, may help understand the underlying mechanism of
these variations.

Rhodes, Reiter, and Schou (2003) used  a few sets of three-day time
series in the computation and fitting of  intermediate-degree {\it
p}-mode frequencies and widths.  The analysis, using Michelson Doppler
Imager (MDI) and Mount Wilson Observatory (MWO) data, resulted in a
higher sensitivity as measured from the slope of the linear regression
between the frequency shifts and activity differences, compared to the
slopes from longer time series.  However, the result was not confirmed
when Global Oscillation Network Group (GONG) data were used (Rose
\etal, 2003).

In the domain of low-degree modes, Chaplin \etal{} (2001) studied
global modes on different time scales, the shortest of which was 27
days. Their results suggest that the sensitivity to changes in the Kitt
Peak magnetic field measurements may be higher in the rising phase of the cycle.
Salabert \etal\ (2002) studied high-frequency modes ($\nu >$ 3.7 mHz) with a time series of eight days using IRIS$^{++}$ data and confirmed that
the frequency shift becomes negative above 4.5~mHz up to the cutoff
frequency of 5.5~mHz. A similar trend for intermediate-degree
frequencies was reported earlier by Ronan, Cadora, and LaBonte (1994)
and Jefferies (1998). 

Here, we study short temporal variations in the frequency shifts and their
degree of correlation with solar activity over a complete solar cycle.
To do so, we use  GONG data for the period 1995\todash{}2005, which
covers partially the descending phase of the cycle 22 and the ascending
and descending part of the current solar cycle 23. The analysis is
carried out by calculating {\it p}-mode frequencies on a time scale as
short as nine days.

\section{Data}

\subsection{Extraction of Mode Frequencies}

We examine time series of nine-days length which were processed with a
multi-taper spectral analysis (Komm \etal, 1999) to produce power
spectra for each spherical harmonic degree (\,$\ell$\,) and order 
(\,$m$\,) up to $\ell$\,=\,100.  The mode frequencies characterized by
$n$, $\ell$, and $m$  were estimated by fitting the individual peaks
(Anderson, Duvall, and Jefferies, 1990).  Each ($n$, $l$) multiplet was then fitted to a  
Legendre polynomial series 
\begin{equation} \nu_{n\ell
m}\,=\,\nu_{n\ell} + \sum_{j=1}^{j_{\rm max}} a_j(n,\ell)P_j(m/\ell) \;,
\end{equation} 
to find the central frequency ($\nu_{n\ell}$) of the
multiplet. Here, $P_j$ is the Legendre polynomial of order $j$, $a_{j}$
are  the splitting coefficients, and  we take $j_{\rm max}$ to be nine.
Since the optimal number of tapers depends on the length of the time series
(Komm \etal, 1999), we use three generalized-sine tapers for the
nine-day time series as opposed to five for the standard GONG procedure
which measures frequencies from time series of 108 days duration. It
should be noted that not all modes are fitted successfully at every
epoch due to the stochastic nature of the modes. More details of the
procedure can be found in Hill \etal{} (1996).

\begin{figure}           
\begin{center}
\psfig{file=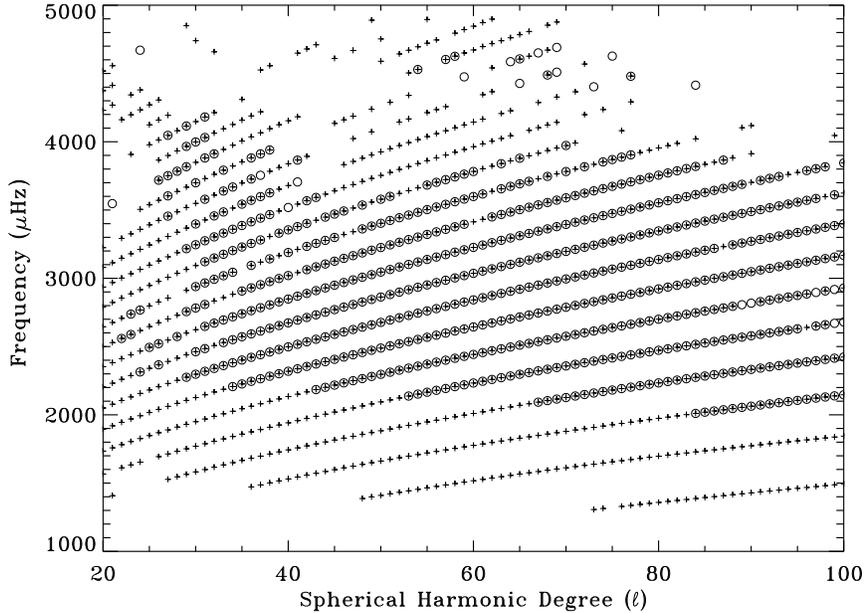,angle=90,width=12cm,clip=}
\caption{The $\ell-\nu$ diagram showing oscillation modes for $\ell \ge
20$.  The circles  denote the modes obtained from nine-day sample
covering the period  16\todash{}24 September 2004 while the pluses
represent the modes from 108-day sample corresponding to the period 8
August to 21 October 2004.}  \label{fig1} \end{center} \end{figure}

The GONG data analyzed here consist of 424 nine-day time intervals  and
cover a period of more than ten years between 7 May 1995 and 16
October 2005.  Each individual nine-day set yields about 650 ($n$,
$\ell$) multiplets in the degree range of   20 $\le \ell \le 100$.  
In Figure~\ref{fig1}, we compare mode frequencies as a
function of $\ell$ obtained from one sample of nine-day and 108-day
(standard GONG product\footnote{\url{
http://gong.nso.edu/data}}) time series.  
It is evident that fewer modes are fitted from the
nine-day power spectra, particularly at low and high frequencies.
Differences between the modes that are present in both data sets are
not visible on this scale.   Further, we notice a few modes
in the nine-day sample that are not present in the 108-day sample,
probably due to their short lifetime.

\subsection{Activity Indices}

We have studied the correlation of the frequencies with five well-known
surface activity indicators observed over the visible surface of the
Sun. These are: the integrated radio flux at 10.7 cm ($F_{10}$)  obtained from
Solar Geophysical Data\footnote{\url{http://www.ngdc.noaa.gov/stp/stp.html}} (SGD), 
the core-to-wing
ratio of the Mg\,{\sc ii} line at 2800~\AA\, (Mg\,{\sc ii})\footnote{
\url{ http://www.sec.noaa.gov/ftpdir/sbuv/NOAAMgII.dat}}, the Mt.
Wilson magnetic plage strength index (\,MPSI\,) and the Mt. Wilson sunspot
index (\,MWSI\,) from Mount Wilson magnetograms (Ulrich, 1991), and the
International Sunspot number ($R_{\rm I}$) obtained from SGD.  These
different measures of solar activity probe the solar atmosphere at
different levels and  show different degrees of correlation with
oscillation frequencies (Bachmann and Brown, 1993; Bhatnagar, Jain, and
Tripathy, 1999).

\begin{figure}           
\begin{center}	\psfig{file=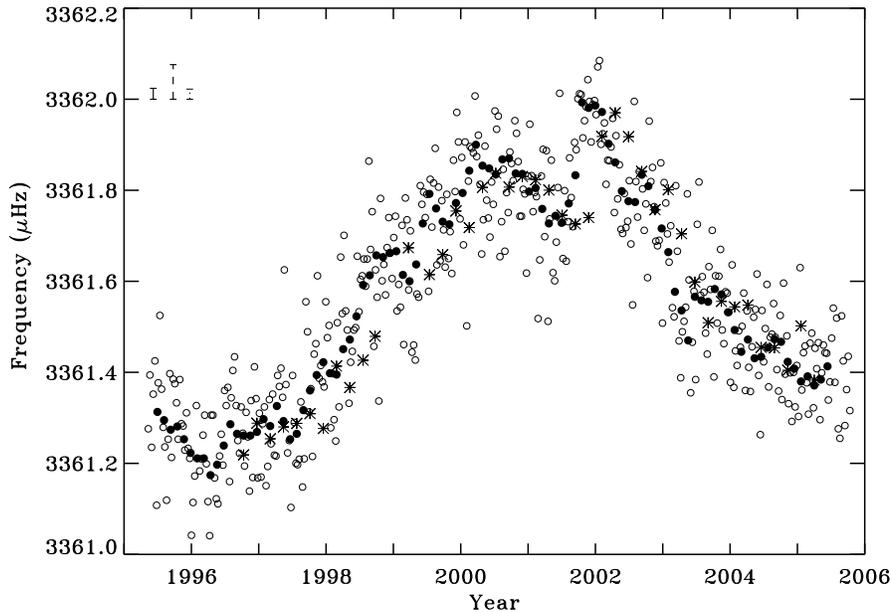,angle=90,width=12.cm,clip=}
\caption{Variation of the central frequency for the $n=9$, $\ell = 81$ mode. 
The open and filled circles represent nine and 108-day time samples from GONG respectively
while the stars represent the 72-day MDI data. 
The lines on the upper-left-hand corner shows representative uncertainties in fitting  
 corresponding to GONG (dashed line for nine days and solid line for 108 days) 
and MDI (dotted) frequencies respectively. }
 \label{fig2}
\end{center}
\end{figure}

\section{Analysis}
As a specific example of the temporal variation of frequency, we show a
single mode corresponding to the $n\,=\,9$, $\ell\,=\,81$ multiplet as
a function of time in Figure~\ref{fig2}. We also include the same mode
from MDI data for corroboration.  A distinct temporal variation can be
seen in all of the data sets.  The change over the solar cycle is about
1\,$\mu$Hz for the nine-day frequencies which is approximately 12 times
larger than the formal uncertainty of 0.075\,$\mu$Hz in the frequency
determination.  The change in the 108-day GONG frequencies and 72-day
MDI frequencies appears to be a smooth average of the nine-day
frequencies.

There are different definitions for calculating
frequency shifts (Howe, Komm, and Hill, 2002).  Here we follow the
approach of Woodard \etal\ (1991) and define the mean frequency
shift (\,$\delta\nu$\,) as the  mode-inertia weighted sum of the measured frequencies:
\begin{equation}
\delta\nu(t)\,=\,{\sum_{n,\ell}\frac{Q_{nl}}{\sigma_{n,\ell}^2}} \delta\nu_{n,\ell}(t)
/\sum_{n,\ell}\frac{Q_{n,l}}{\sigma_{n,\ell}^2} , 
\end{equation}
where
$\sigma_{n,\ell}$ is the uncertainty in the frequency measurement,
$\delta\nu_{n,\ell}(t)$ is the change in a given multiplet of $n$ and
$\ell$, and $Q_{n,\ell}$ is the inertia ratio as defined by
Christensen-Dalsgaard and Berthomieu (1991). The reference frequency is
chosen in such a way that it corresponds to a minimum value of the
activity; the 36th set covering the period 17\todash{}25 March
1996 was used. 

In order to investigate the variation  of the frequencies with solar
activity, we correlate and fit the shifts against the activity
differences using the expression \be \delta\nu\ =\ a\; \delta I\ +\ b
\ee to give the best linear fit to the data. The slope (\,$a$\,) which
measures the shift per activity index and hence sensitivity of the
shift, and  the intercept (\,$b$\,) are obtained from the linear
least-square fit for each of the activity index (\,$I$\,). Following the
definition of frequency shift,  $\delta I$ is calculated by taking the
36th data point as the reference activity unless mentioned otherwise in
the text.

\begin{figure}[h]          
\begin{center}
\psfig{file=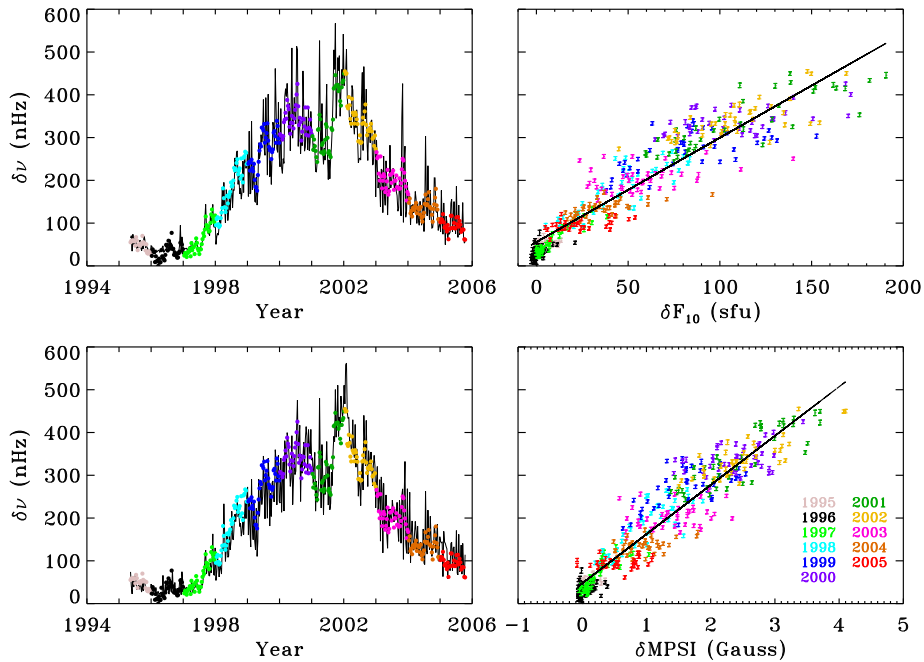,angle=90,width=\linewidth} \caption{Mean-frequency
shifts plotted as a function of both time (left panels) and change in
activity (right panels). The solid lines  in the left-hand panels are the
scaled frequency differences which result from the two least-squares fits which are shown in the 
right-hand panels.  The solid lines in those two panels
denote the linear fits to the frequency differences when regressed upon changes in the 10.7-cm radio flux 
and upon changes in MPSI. The
corresponding slopes and intercepts are given in the last rows of Tables~1
and 2. For clarity, the 1$\sigma$ uncertainties in the mean
frequencies are omitted from the left panels. These are visible in the
right panels. }
\label{fig3}
\end{center}
\end{figure}

\section{Results and Discussions}
Figure~\ref{fig3} shows the mean frequency shifts as a function of the
epoch (left panel) and the change in activity (right panel) for two
activity indices; one magnetic (MPSI) and the other non-magnetic
($F_{10}$). It is reassuring
to see that the mean frequency shifts derived from nine-days clearly
show a similar solar cycle variation  to those derived from longer time
series. The only difference is the rapid fluctuations seen in each of
the plots which arises due to the short interval over which the mean
frequency shifts and mean activity differences are calculated.
The solid lines in the two right-hand panels represent the straight lines
 obtained by using Equation~(3) to give the best linear fits to the two sets of data points;
 the corresponding slopes and intercepts of both fits are tabulated in the
last rows of Tables~1 and 2.  A significant correlation between the 
frequency shifts  and change in activity indices is evident except near
the activity-minimum phase of the cycle around 1996\todash{}97
reflecting deviations from the assumed linear dependence.  We examine
this more critically in the next Section.  

\subsection{Year-wise Variations} 
\begin{figure}             
\begin{center}
\psfig{file=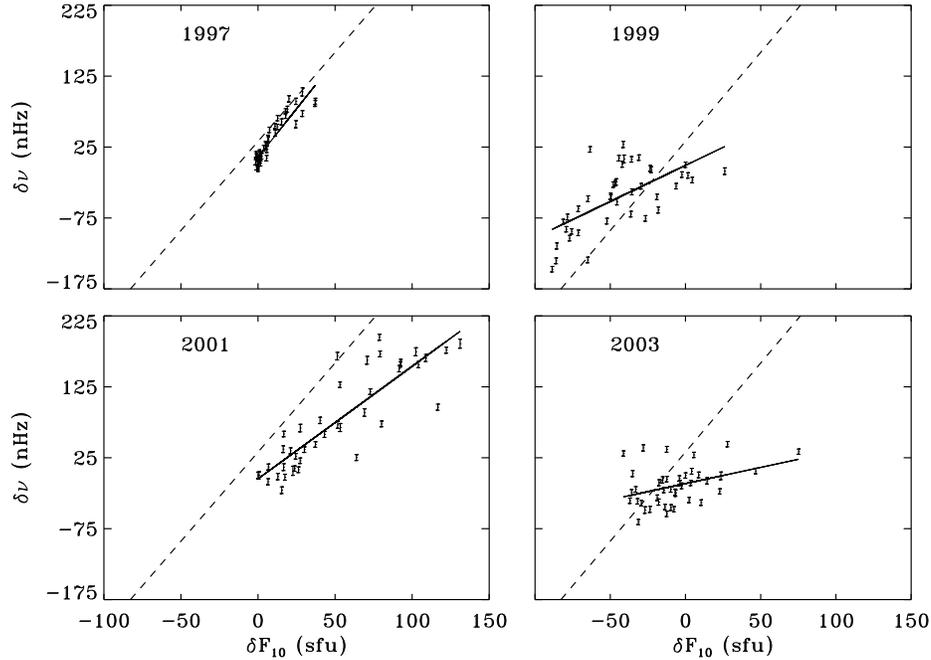,angle=90,width=\linewidth} 
\caption{Mean-frequency shifts plotted as a function of  $\delta F_{10}$ for four
selected years as marked on each figure.  The solid line denotes the
linear fit to the activity measurements for that year. The error bars
signify the 1$\sigma$ uncertainties in the mean-frequencies.  The
dashed line denotes the linear fit to all 424 data points.}
\label{fig4} 
\end{center} 
\end{figure}

\begin{figure}              
\begin{center}
\psfig{file=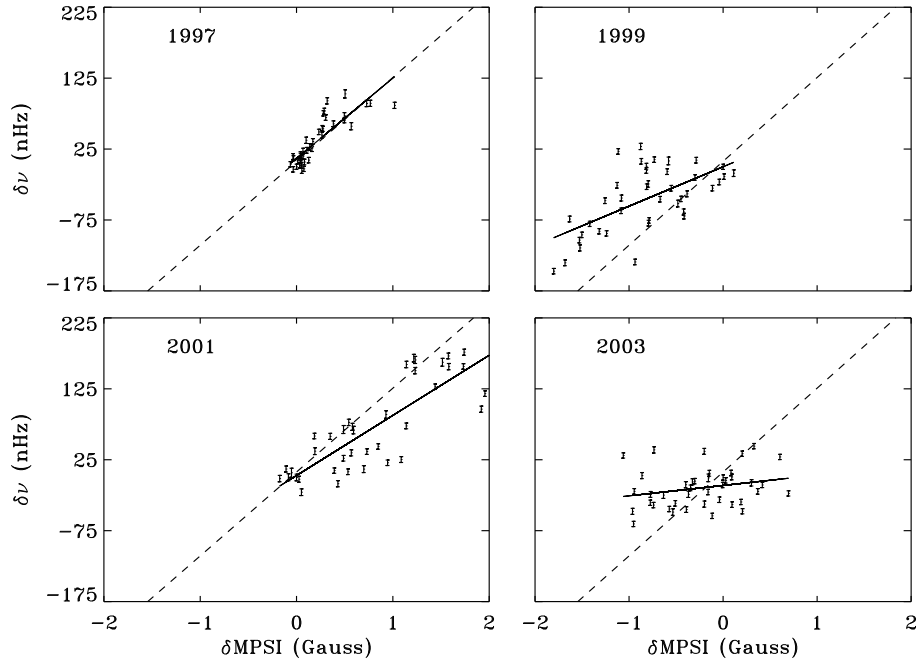,angle=90,width=\linewidth}            
\caption{Same as Figure~\ref{fig4} but for $\delta$MPSI}.
\label{fig5}
\end{center}
\end{figure}

\begin{figure}          
\begin{center}
\vskip -1.5in
\psfig{file=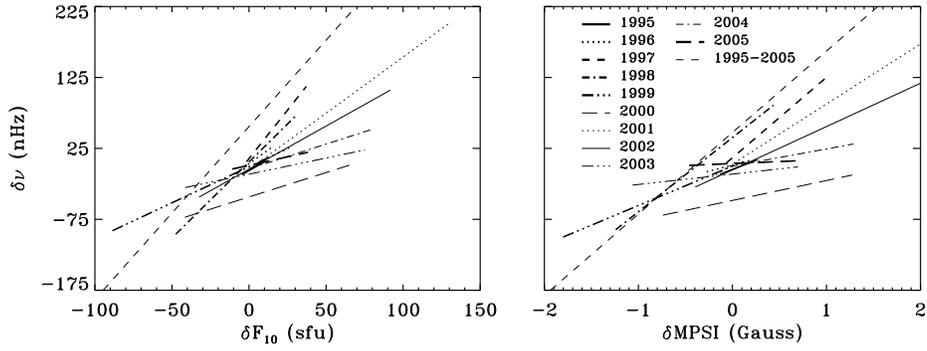,angle=90,width=\linewidth}
\caption{The panels shows the linear fit 
between the year-wise frequency shifts and activity differences and
are constructed using the slopes and intercept values  given in Table~1 
(left panel) and Table~2 (right panel). }
\label{fig6}
\end{center}
\end{figure}

In order to study how closely the frequencies follow the
activity levels, we study year-wise variations. The distribution of the
number of points in each year can be seen in Table~1. As before, the
mean frequency shift for each year (also for the entire data set) is
calculated using Equation~(2) but the reference frequency is now
considered to be the middle point in each set.  Similarly, the middle
point in each activity set was considered as the reference activity for
the linear regression analysis. For example, for the year 1995, the
reference point for calculating the frequency shift and change in
activity was taken as the 14th point.  The results of the linear
least-square fits and correlation coefficients for $F_{10}$ are
summarized in Table~1. For comparison, we also tabulate the slope and
correlation coefficient obtained from the entire data set in the last
row of the Table~1. We find significant year-to-year variation in the
calculated slopes and the correlation coefficients.  For a more
detailed investigation of the variation in the slopes, we show mean-frequency 
shifts as a function of the activity differences 
($\delta F_{10}$)  for four  representative  years in
Figure~\ref{fig4}.  The solid lines in the four panels represent the linear fits to the pairs of 
frequency differences and 10.7-cm flux differences in the four different years, while the dashed line which is repeated 
in all four panels, is the linear fit to all 424 datasets. This linear fit can be considered  as 
the representative fit for the current solar cycle.
It is  interesting to note that the solid and dashed lines for the year
1997 have similar orientation implying close agreement in the values of
the slopes indicating similar mean shifts per unit activity. Surprisingly, 
the best correlation between the frequency shifts  and
10.7 cm radio flux also exists during 1997 which 
represents the
beginning of the rising phase of the current solar cycle.  Results for
MPSI obtained with the similar analysis  are given in Table~2 and
Figure~\ref{fig5}.  For MPSI, a significant correlation is obtained
for  1997, 1998, and 2001 and a poor correlation for  1996, 2003, and
2005, and it closely agrees with $F_{10}$.  Similar results are found for
all other activity indices considered in this paper.  Specifically, we
note that the correlation coefficients and slopes obtained from the
analysis of the entire data sets are similar to those of 1997 or slightly
higher and do not represent average values of the year-wise
variations.  This can be more clearly visualized in
Figure~\ref{fig6} where we have shown the linear fits between the
frequency shifts and activity differences using the slopes and
intercepts  given in Tables~1 and 2. Numerically, the average
slopes of the yearly regression fits are 1.26 nHz/sfu for $F_{10}$ and 52.57 nHz/Gauss
for MPSI. These mean slopes are each approximately a  factor of
two smaller than are the slopes of the overall fits to the corresponding collection of 424 datasets.


In a similar analysis, Woodard \etal\ (1991) have studied the variation
of oscillation frequencies during 1986, 1988, and 1989. They report
different correlation coefficients for annual variations (except for
the magnetic field during 1988 and 1989) which are also different when
all the  data were taken together. Our result qualitatively agrees with
these findings although there are differences in detail. However, the
discrepancy can not be investigated further since the two studies do
not have common activity indices.  

\begin{figure}[h]           
\begin{center}
\psfig{file=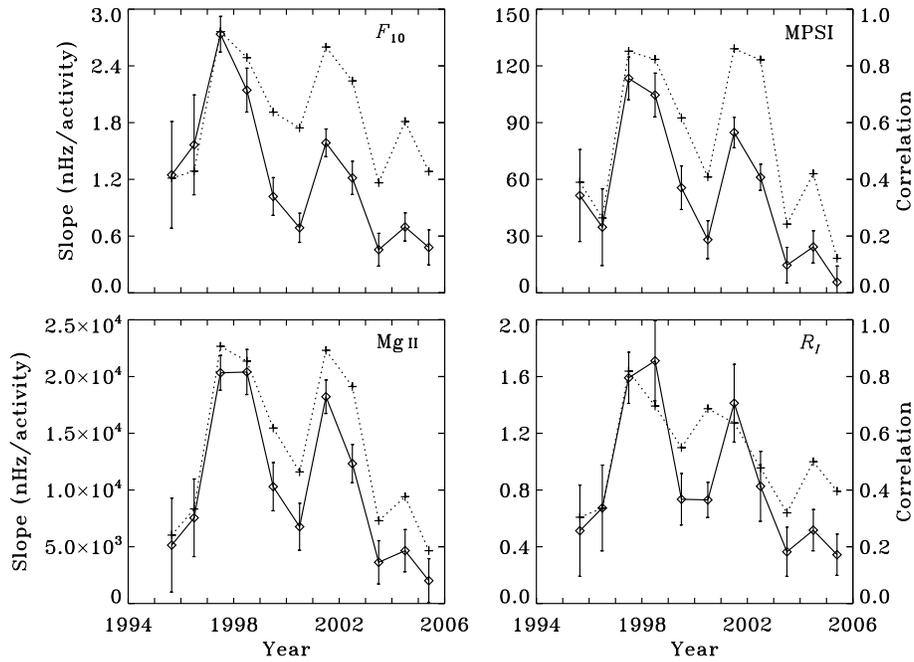,angle=90,width=\linewidth}
\caption{Year-wise variation of the linear-regression slope (diamonds)
and the Pearson's correlation coefficients (pluses) for different activity indices.
A strong correlation between the two is clearly visible. The error bars  
signify the $1\sigma$ errors in the determination of the slopes.}\label{fig7}
\end{center}
\end{figure}

Since the linear regression analysis
clearly shows the  yearly variations in the slopes and the correlation
coefficients, we investigate a possible correlation between these two
quantities. This is carried out by plotting the annual variation of the correlation
coefficients along with the slopes for four out of the five activity
indicators considered in this paper (Figure ~\ref{fig7}). The
similarity between the two curves stands out in each case signifying
strong correlations, although some differences can be seen in detail.
The correlation is further confirmed by calculating the linear and rank
correlation coefficients between the two quantities (Table~3). The
maximum correlation is obtained for \ion{Mg}{II}, followed by MPSI
while MWSI has the lowest correlation coefficient. 

\begin{figure}           
\begin{center}
\psfig{file=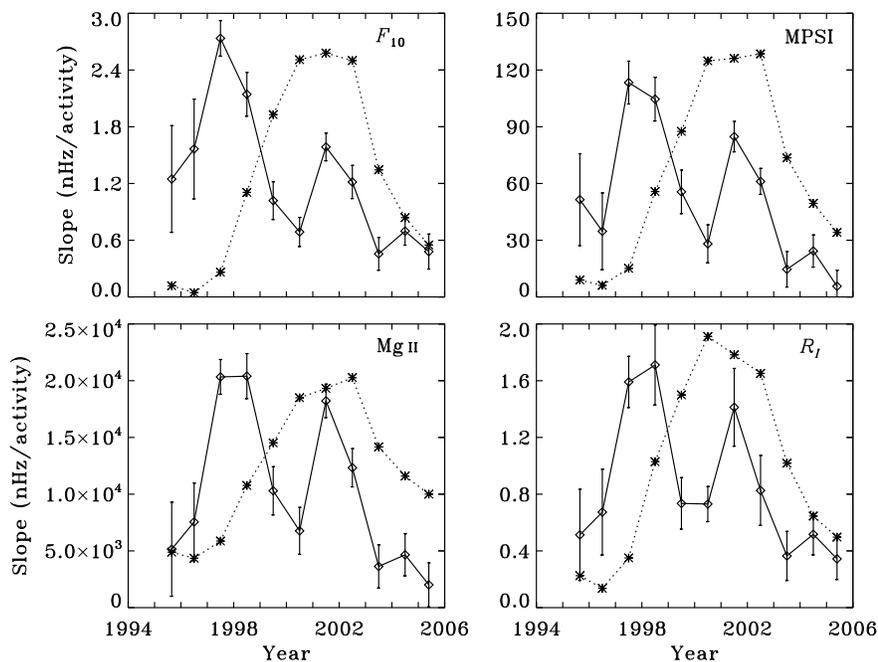,angle=90,width=\linewidth}
\caption{Temporal dependence of the linear-regression slopes (diamonds)
and corresponding  yearly-averaged solar activity (stars). The error bars 
signify the 1$\sigma$ errors in the determination of the slopes.} 
\label{fig8}
\end{center}
\end{figure}

The relationship between the annual linear-regression slopes and yearly
averaged solar activity is shown in Figure~\ref{fig8}.  The striking
feature is the strong temporal variability of the regression slopes
which do not appear to be correlated with activity indices on the time
scale of the solar cycle.  On closer examination, there appears to be a good
correlation between 1995--1998 and 2001--2005.  This could indicate
that there are two components of the activity which contribute to the
variability of the slopes.  The major contribution comes from the
overall weak magnetic field, and hence during the ascending and
descending phase  of the cycle we have a strong correlation with
activity.  But during the maximum phase of the cycle, when the field is
dominated by the strong field, due to the increase in the number and
strength of the active regions, the slopes do not follow the activity
measurements.

\begin{landscape}
\begin{center}       
\begin{table}
\caption{Year-wise distribution of correlation statistics between  frequency shifts (\,$\delta\nu$\,) and 
$\delta F_{10}$. Shown are the number of data sets (\,$N$\,), epoch covered, 
the mean shift per unit change in activity (\,$a$\,) in nHz/sfu, the intercept
 (\,$b$\,) in nHz, 
Pearson's linear coefficient (\,$P_{\rm p}$\,)
Spearman rank correlation coefficient (\,$r_{\rm s}$\,) and its two-sided significance 
(\,$P_{\rm s}$\,). The last row denotes the values corresponding to all
424 data sets. }
\begin{tabular}{rr@{ \todash {}}rr@{ $ \pm $ }rr@{ $ \pm $ }rccl}
\hline
$N$&\multicolumn{2}{c}{Epoch}&\multicolumn{2}{c}{$a$}&\multicolumn{2}{c}{$b$}&$P_{\rm p}$&$r_{\rm s}$&$P_{\rm s}$\\
\hline\noalign{\smallskip}
27&7 May 1995 & 4 January 1996&1.25 & 0.56&-6.33&2.70&0.40&0.40&3.9 $\times\ 10^{-2}$\\
41&5 January 1996&  7 January 1997&1.56 & 0.53&3.42&3.45&0.42&0.42&5.9 $\times\ 10^{-3}$\\
40&8 January 1997&  2 January 1998&2.73 & 0.19&10.62&2.68&0.92&0.92&6.1 $\times\ 10^{-17}$\\
41&3 January 1998&  6 January 1999&2.14 & 0.23&5.88&5.06&0.83&0.79&1.1 $\times\ 10^{-9}$\\
40&7 January 1999&  1 January 2000&1.02 & 0.20&-0.83&10.40&0.64&0.64&9.4 $\times\ 10^{-6}$\\
41&2 January 2000&  4 January 2001&0.69 & 0.15&-43.47&4.17&0.58&0.50&9.8 $\times\ 10^{-4}$\\
41&5 January 2001&  8 January 2002&1.59 & 0.15&-4.74&9.30&0.87&0.88&6.4 $\times\ 10^{-14}$\\
40&9 January 2002&  3 January 2003&1.22 & 0.18&-4.32&7.09&0.75&0.76&1.5 $\times\ 10^{-8}$\\
41&4 January 2003&  7 January 2004&0.46 & 0.17&-11.38&4.21&0.39&0.37&1.9 $\times\ 10^{-2}$\\
40&8 January 2004&  1 January 2005&0.70 & 0.15&-3.75&3.80&0.60&0.66&3.4 $\times\ 10^{-6}$\\
32&2 January 2005&  16 October 2005&0.48 & 0.18&1.13&2.79&0.43&0.33&6.5 $\times\ 10^{-2}$\\
424&7  May 1995&  16 October 2005&2.51 & 0.04&32.98&5.54&0.94&0.95&0.0\\
\noalign{\smallskip}\hline
\end{tabular}
\end{table}
\end{center}
\end{landscape}
\begin{landscape}
\begin{center}      
\begin{table}
\caption{Year-wise distribution of correlation statistics between frequency shifts (\,$\delta\nu$\,) and 
$\delta{\rm{MPSI}}$. The notations have the same meaning as in Table~1. The
units of $a$ are nHz/Gauss.}
\begin{tabular}{rr@{ \todash {}}rr@{ $ \pm $ }rr@{ $ \pm $ }rccl}
\hline
$N$&\multicolumn{2}{c}{Epoch}&\multicolumn{2}{c}{$a$}&\multicolumn{2}{c}{$b$}&$P_{\rm p}$&$r_{\rm s}$&$P_{\rm s}$\\
\hline\noalign{\smallskip}
27&7 May 1995 & 4 January 1996 &51.45 & 24.30&-4.65&2.53&0.39&0.36&6.2 $\times\ 10^{-2}$\\
41&5 January 1996 & 7 January 1997&34.71 & 20.30&1.49&4.15&0.26&0.24&1.3 $\times\ 10^{-1}$\\
40&8 January 1997 & 2 January 1998&113.35 & 11.31&11.79&3.64&0.85&0.87&2.1 $\times\ 10^{-13}$\\
41&3 January 1998 & 6 January 1999&104.59 & 11.56&40.20&7.09&0.82&0.80&3.4 $\times\ 10^{-10}$\\
40&7 January 1999 & 1 January 2000&55.57& 11.51&0.06&11.04&0.61&0.54&3.5 $\times\ 10^{-4}$\\
41&2 January 2000 & 4 January 2001&28.10 &  10.05&-48.39&5.16&0.41&0.32&4.0 $\times\ 10^{-2}$\\
41&5 January 2001 & 8 January 2002&84.78 &  8.04&2.79&8.96&0.86&0.86&5.3 $\times\ 10^{-13}$\\
40&9 January 2002 & 3 January 2003&61.10&  6.88&-5.32&5.86&0.82&0.78&2.3 $\times\ 10^{-9}$\\
41&4 January 2003 & 7 January 2004&14.63 &  9.39&-11.23&4.77&0.24&0.27&8.9 $\times\ 10^{-2}$\\
40&8 January 2004 & 1 January 2005&24.29 &  8.52&-0.06&4.47&0.42&0.36&2.4 $\times\ 10^{-2}$\\
32&2 January 2005 & 16 October 2005&5.68 & 8.46&3.65&3.10&0.12&0.11&5.5 $\times\ 10^{-1}$\\
424&7  May 1995 & 16 October 2005 &118.25 & 2.10&7.70&5.03&0.94&0.94&0.0\\
\noalign{\smallskip}\hline
\end{tabular}
\end{table}
\end{center}
\end{landscape}

\begin{center} 
\begin{table}
\caption{The correlation statistics between slopes and correlation coefficients for different 
activity indicators. Shown are the Pearson's linear coefficient (\,$P_{\rm p}$\,), the rank correlation (\,$r_{\rm s}$\,), and 
the two-sided significance (\,$P_{\rm s}$\,).}
\begin{tabular*}{0.60\textwidth}%
{@{\extracolsep{\fill}}rccr}
\hline\noalign{\smallskip}
\multicolumn{1}{c}{Activity}&$P_{\rm p}$&$r_{\rm s}$&\multicolumn{1}{c}{$P_{\rm s}$}\\
\hline
$F_{10}$ &0.72&0.72&1.1 $\times$ $10^{-2}$\\
MPSI&0.90&0.89&2.3 $\times$ $10^{-4}$\\
MWSI&0.57&0.58&6.0 $\times$ $10^{-2}$\\
Mg\,{\sc ii}&0.96&0.91&1.1 $\times$ $10^{-4}$\\ 
$R_{\rm I}$&0.81&0.81&2.6 $\times$ $10^{-3}$\\
\noalign{\smallskip}\hline
\end{tabular*}
\end{table}
\end{center}

We finally note that the year-wise variation of the oscillation
frequencies is distinctly different than the analysis involving the
entire data sets emphasizing that both the long and short--term
variations are different in nature. We also find that the long-term
variation is not a simple average of the short-term variations.
Therefore, we conclude that studies of changes of oscillation
frequencies  which consider all the data sets together may not
accurately represent the variations on a scale which is smaller than
the length of the entire data sets.

\subsection{Variation in Rising and Falling Phases of Activity Cycle 23}
From the discussion given in Section~4.1, it is evident that there is
no significant difference in the correlation pattern between the
magnetic (Table~2) and non-magnetic (Table~1) activity indices.  But
the analysis  of Chaplin \etal\ (2001) involving low-degree modes
suggested that the sensitivity to changes in the disk-averaged
line-of-sight magnetic field component may be higher during the rising
phase of the cycle.  We perform a similar analysis by separating our
data set into rising and falling phases of activity cycle 23. Since May
1996 and April 2000 mark the minimum and maximum of cycle 23 (see SGD),
we choose the period 1 May 1996 \todash{}15 May 2000  consisting of
1476 days  as the rising phase.  Avoiding the secondary maximum of this
cycle,  we choose 02 October 2001 \todash{}16 October 2005  again
consisting of 1476 days as the falling phase so that both halves have a
similar minimum to maximum range in activity indices. Further, to be
consistent with the analysis of annual variations, $\delta I$ is
calculated by taking the middle data point in each set as the reference. 
The results are summarized in Table~4 where we have shown the
slopes, the rank correlation coefficients, and the differences in the
slopes  between descending and ascending phases
normalized by their  combined uncertainty ($\sigma_{\delta a}$); a negative $\sigma_{\delta
a}$  indicating a higher value for the rising phase.  The differences for magnetic
indices, MPSI and MWSI, with $\sigma_{\delta a}~>$~4 suggest
significant differences in behaviour of the magnetic flux emerging over
the ascending and descending phases.  The value of 2.79 for the
10.7 cm radio flux implies behaviour similar to the magnetic indices
but not as significant as the magnetic flux, while {Mg}{\sc II} with a
value closer to zero implies no phase dependence.  The sunspot number,
the only index with a positive $\sigma_{\delta a}$  exhibits an
opposite behaviour compared to other indices, but since its value 
is less than 2, the difference can not be considered to be of
significance.  This result broadly agrees with the findings of Chaplin
\etal{} (2001) where a phase-dependent behaviour   for the Kitt Peak
magnetic index was reported; but it is not clear why other magnetic
indices {\it e.g.} MPSI and MWSI did not show a similar difference.
Our results, on the other hand, shows phase-dependent behaviour for all
of the magnetic indices and  suggest that the frequency shifts are
caused by the magnetic fields at or very near the solar surface.

Similarly, the rank correlations obtained between the frequency shifts and activity indicators 
also manifest a phase-dependent pattern. In general, the rank correlation coefficients are 
lower in the descending phase of the activity cycle.  
The maximum change is seen in the case of MWSI but we should point out that
there are significant gaps in the measurement of MWSI index which may
contribute to this discrepancy.  Finally, we speculate that this
phase-dependent behaviour of the frequency shifts with magnetic flux
gives rise to the  hysteresis pattern seen only in case of the magnetic
activity indices (Jim{\'e}nez-Reyes \etal, 1998 and  Tripathy \etal,
2001).

\begin{landscape}
\begin{center}     
\begin{table}
\caption{Results of linear fits to the change in activity indices in the rising and falling phases
of cycle 23. Each phase consists of about four years of data. Shown are
the mean shift per unit change in activity (\,$a$\,), the intercept (\,$b$\,), 
Spearman rank correlation coefficient (\,$r_{\rm s}$\,).  The last column (\,$\sigma_{\delta a}$\,) gives 
the difference in the slopes between ascending and descending phases normalized 
by the combined uncertainty. We have omitted the two-sided significance 
as these are all zeros indicating high significance except for
MWSI  with a value of $10^{-22}$ in the descending phase.}
\begin{tabular}{lr@{ $\pm$ }rr@{ $\pm$ }rcr@{ $\pm$ }rr@{ $\pm$ }rcr}
\hline\noalign{\smallskip}
&\multicolumn{5}{c}{Ascending phase}&\multicolumn{5}{c}{Descending phase}&\\
{Activity}&\multicolumn{2}{c}{$a$}&\multicolumn{2}{c}{$b$}&$r_{\rm s}$&\multicolumn{2}{c}{$a$}&\multicolumn{2}{c}{$b$}
&$r_{\rm s}$&$\sigma_{\delta a}$\\
index&\multicolumn{2}{c}{(nHz/index)}&\multicolumn{2}{c}{(nHz)}&&\multicolumn{2}{c}{(nHz/index)}&\multicolumn{2}{c}{(nHz)}&\\
\hline\noalign{\smallskip}
$F_{10}$&2.54 & 0.08&180.07 & 3.12&0.95&2.32 & 0.08&174.44 & 3.66&0.90&$-$2.79\\
{MPSI}&129.72 & 3.67&142.63 & 3.01&0.94&109.08 & 3.54&202.59 & 3.28&0.88&$-$5.72\\
{MWSI}&358.29 & 26.12&167.21 & 5.93&0.86&248.41& 24.48&184.73 & 7.23&0.67&$-$4.34\\
{Mg\,{\sc ii}}&21476~~~~&516~~~~ &162.86 & 2.55&0.95&21326~~~~&638~~~ &199.47 & 3.07&0.90&$-$0.26\\
$R_I$&2.30 & 0.08&190.29 & 3.82&0.92&2.49 & 0.12&184.51 & 4.64&0.84&1.84 \\
\noalign{\smallskip}\hline
\end{tabular}
\end{table}
\end{center}
\end{landscape}

\begin{center}     
\begin{table}[ht]
\caption{Results of linear fits to differences in activity indices and correlation statistics 
for  frequency shifts obtained from time series of different length. 
Shown are the length of the time (\,$t$\,),
the mean shift per unit change in activity (\,$a$\,), 
 the intercept (\,$b$\,), Pearson's linear coefficient (\,$P_{\rm p}$\,), Spearman rank correlation (\,$r_{\rm s}$\,). The two-sided 
significance (\,$P_{\rm s}$\,) is smaller than $10^{-17}$ and is not shown.}
\begin{tabular*}{0.80\textwidth}%
{@{\extracolsep{\fill}}crrrcc}
\hline\noalign{\smallskip}
Activity&\multicolumn{1}{c}{$t$}&\multicolumn{1}{c}{$a$} &\multicolumn{1}{c}{$b$}&$P_{\rm p}$&$r_{\rm s}$\\
index&\multicolumn{1}{c}{(days)}&\multicolumn{1}{c}{(nHz/index)}&\multicolumn{1}{c}{(nHz)}&&\\
\hline
$F_{10}$&9 &2.44 $\pm$ 0.04&55.72 $\pm$ 3.24&0.94&0.95\\
&72&1.75 $\pm$ 0.04&15.37 $\pm$ 2.69&0.99&0.99\\
&108&1.77 $\pm$ 0.02&14.57 $\pm$ 1.82&0.99&0.99\\
\hline
MPSI&9&114.90 $\pm$ 2.05&47.41 $\pm$ 3.27& 0.94&0.94\\
&72&79.45 $\pm$ 2.23&15.49 $\pm$ 3.57&0.98&0.98\\
&108&80.41 $\pm$ 1.34&11.93 $\pm$ 2.14&0.99&0.98\\
\hline
MWSI&9&427.6 $\pm$ 22.7&123.93 $\pm$ 5.63 &0.68&0.80\\
&72&481.7 $\pm$ 40.5&34.94 $\pm$ 8.96&0.87&0.89\\
&108&504.6 $\pm$ 25.9&29.57 $\pm$ 5.65&0.89&0.91\\
\hline
{Mg\,{\sc ii}}&9&19621~ $\pm$ 344~~&23.38 $\pm$ 3.56&0.94&0.94\\
&72&13601~ $\pm$ 494~~&0.26 $\pm$ 5.07&0.97&0.97\\
&108&13330~ $\pm$ 312~~&1.47 $\pm$ 3.20&0.97&0.97\\
\hline
$R_{\rm I}$&9&2.40 $\pm$ 0.06&65.77 $\pm$ 3.96&0.90&0.92\\
&72&1.81 $\pm$ 0.07&12.80 $\pm$ 5.21&0.97&0.95\\
&108&1.85 $\pm$ 0.04&8.72 $\pm$ 2.97&0.97&0.97\\
\noalign{\smallskip}\hline
\end{tabular*}
\end{table}
\end{center}

\subsection{Comparison With Standard GONG and MDI Data}
It is expected that with longer time series, the determination of the
frequencies will be more precise as the uncertainty is inversely
proportional to the square root of the length of the time series for
resolved modes -- those with lifetimes less than the length of the time
series (Libbrecht, 1992).  Additionally, small fluctuations in activity
would be averaged out during the longer time period. Thus, it appears
that the correlation between the frequency shift and activity will
depend on the length of the observing run.  The analysis involving
low-degree modes (Chaplin \etal, 2001) on different time scales, from
216 days to 27 days, confirms that variations in rank correlation are
not inconsistent with the change in accuracy expected for different
observing intervals. They also report a small rank-correlation
coefficient of the order of 0.67 on the time scale of 27 days which is
independent of all of the activity indices.  If we extrapolate this
result to a temporal interval of nine days, we expect to find  a very
small correlation between the activity indices and frequency shifts.

We compare our results obtained from three different time samples, nine
days and 108 days from GONG, and 72 days from MDI in Table~5. The 108-day 
GONG data covering the period between 7 May 1995 and 16 October 2005
consists of 104 sets with an overlapping period of 36 days between the
two adjacent data sets. The MDI data consists of 46 intervals spanning
the period from 1 May 1996 to  16 October 2005. 
The reference frequency for each data set is
chosen in such a way that it corresponds to a minimum value of the
activity. 
From Table~5, we find that all three correlations are highly significant
with slight variations in the degree of correlation probably due to the
different length of the time series.  For the nine-day frequencies, we
find the smallest correlation coefficients but they are significantly
higher than the values corresponding to analysis of low-degree modes on
time scale of 27 days (Chaplin \etal, 2001).  Since 
localized  active regions have much more power at high degrees,  the low and 
intermediate-degree modes have different responses to activity indices. We 
also note that for a data set of given length, no significant
difference  is found between correlations among different activity
indices except in the case of MWSI. This is probably due to the large
number of gaps in the data which resulted in a poor average when the
average was taken over nine days and had no significant effect for
averages taken over longer periods.

Similar evidence for different behaviour between low and intermediate-degree 
modes comes from the comparison of the sensitivity values. 
We find a higher sensitivity for the short-duration frequencies and
confirm the findings of Rhodes, Reiter, and Schou (2003). However, we
do not find any explanation for the results of Rose \etal{} (2003) who
could not confirm the higher value of sensitivity using GONG$^{++}$ data.
It is possible that during the high-activity period of 2001, the
variation in sensitivity is too small to be detected. 
This result is again different from that of Chaplin \etal{} (2001),
who found similar values for all time intervals considered in their study.
Finally, it is worth noting that only in the case of the MWSI index, the slope 
from the nine-day time series is smaller than the slopes from either the 72- or 108-day
time series and may again be attributed to the problem of data gaps as discussed earlier.

\begin{figure}           
\begin{center}
\psfig{file=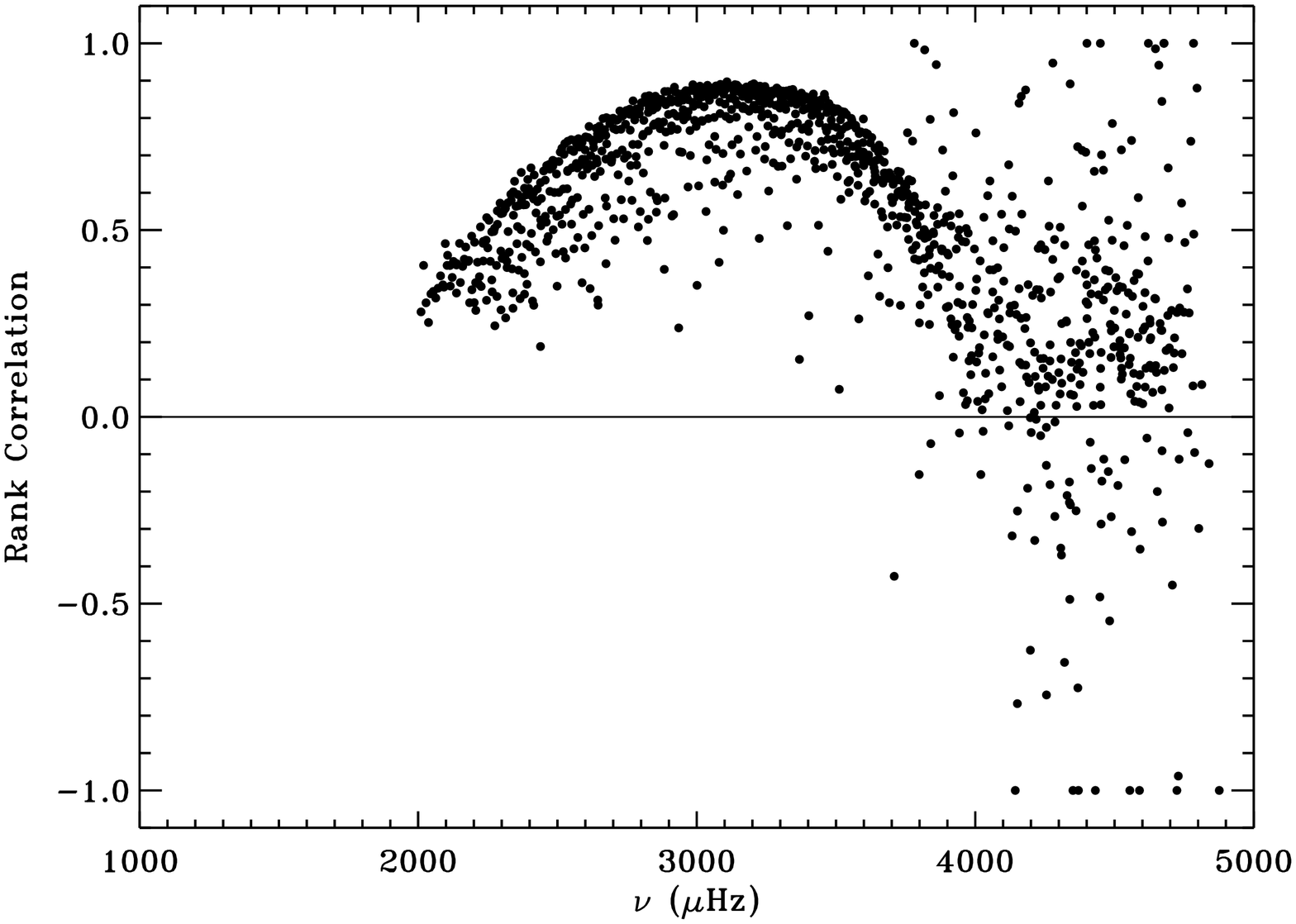,angle=90,width=12.cm,clip=}
\psfig{file=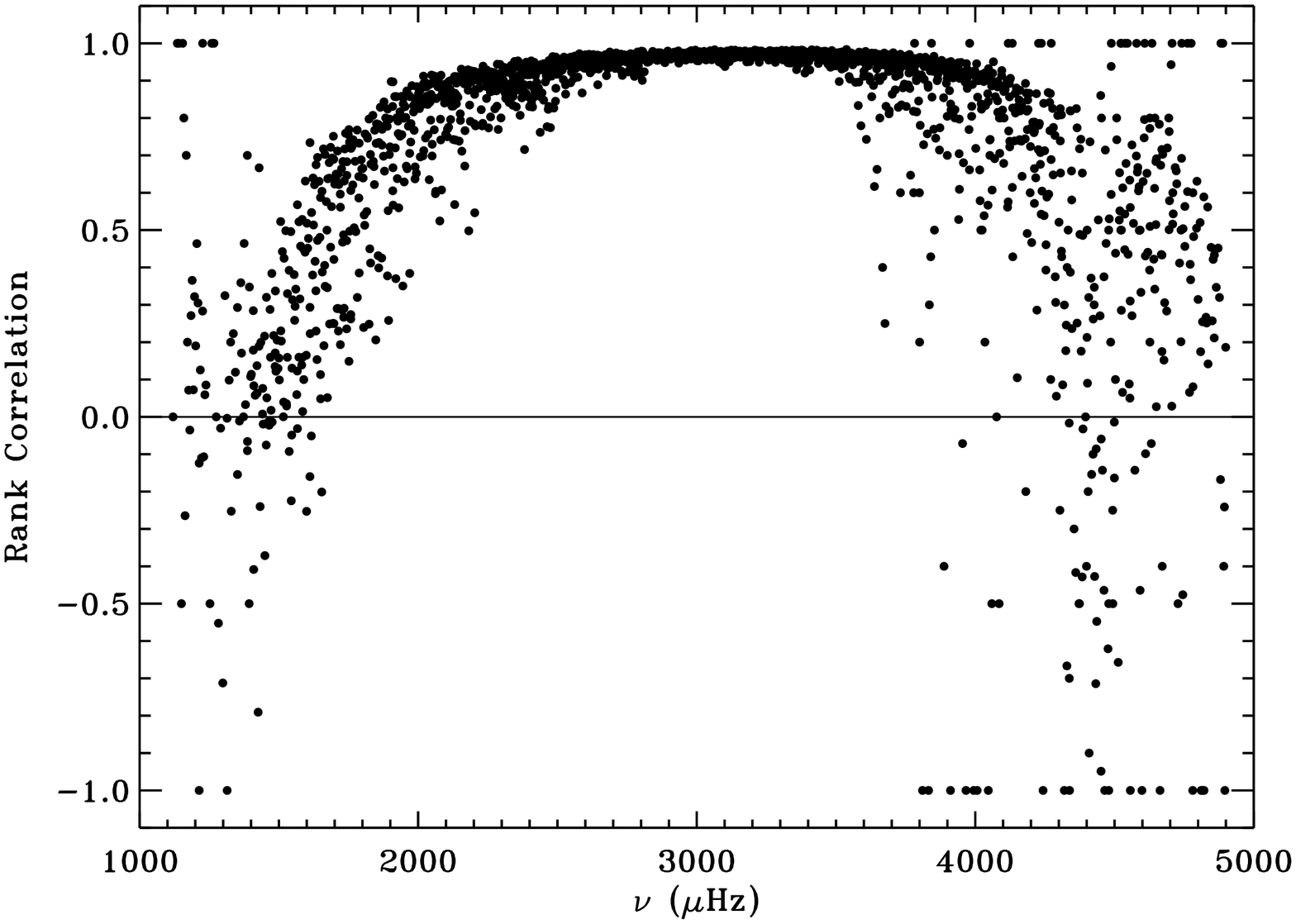,angle=90,width=12.cm,clip=}
\caption{The Spearman rank correlation coefficients, as a function of frequency, between  
the weighted frequency shifts and $\delta{\rm{MPSI}}$ for all modes present in
nine-day
(top panel) and 108-day frequency determinations (bottom panel).}   
\label{indicor}
\end{center}
\end{figure}
\begin{figure}  
\begin{center}
\psfig{file=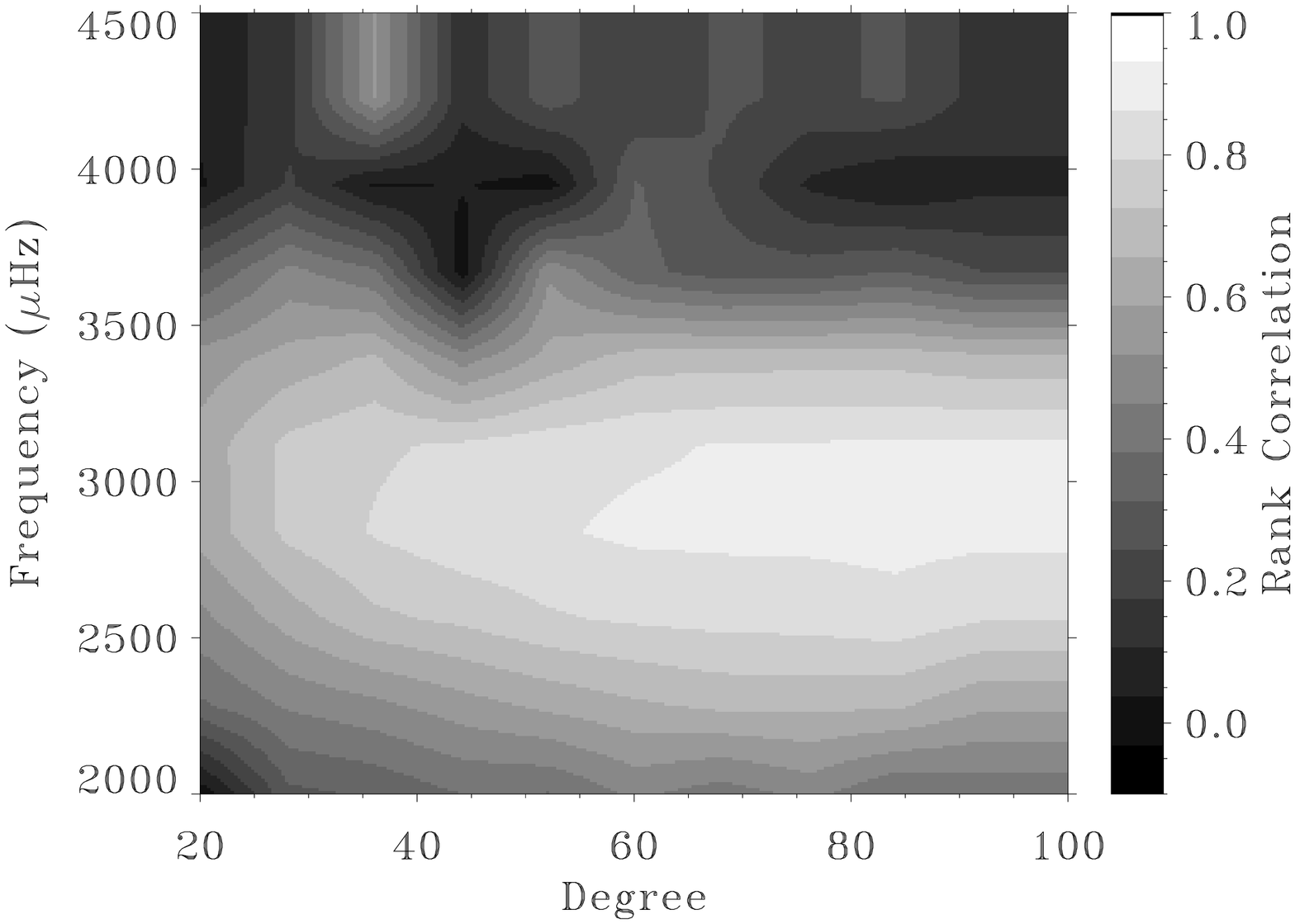,angle=90,width=12.cm,clip=}
\psfig{file=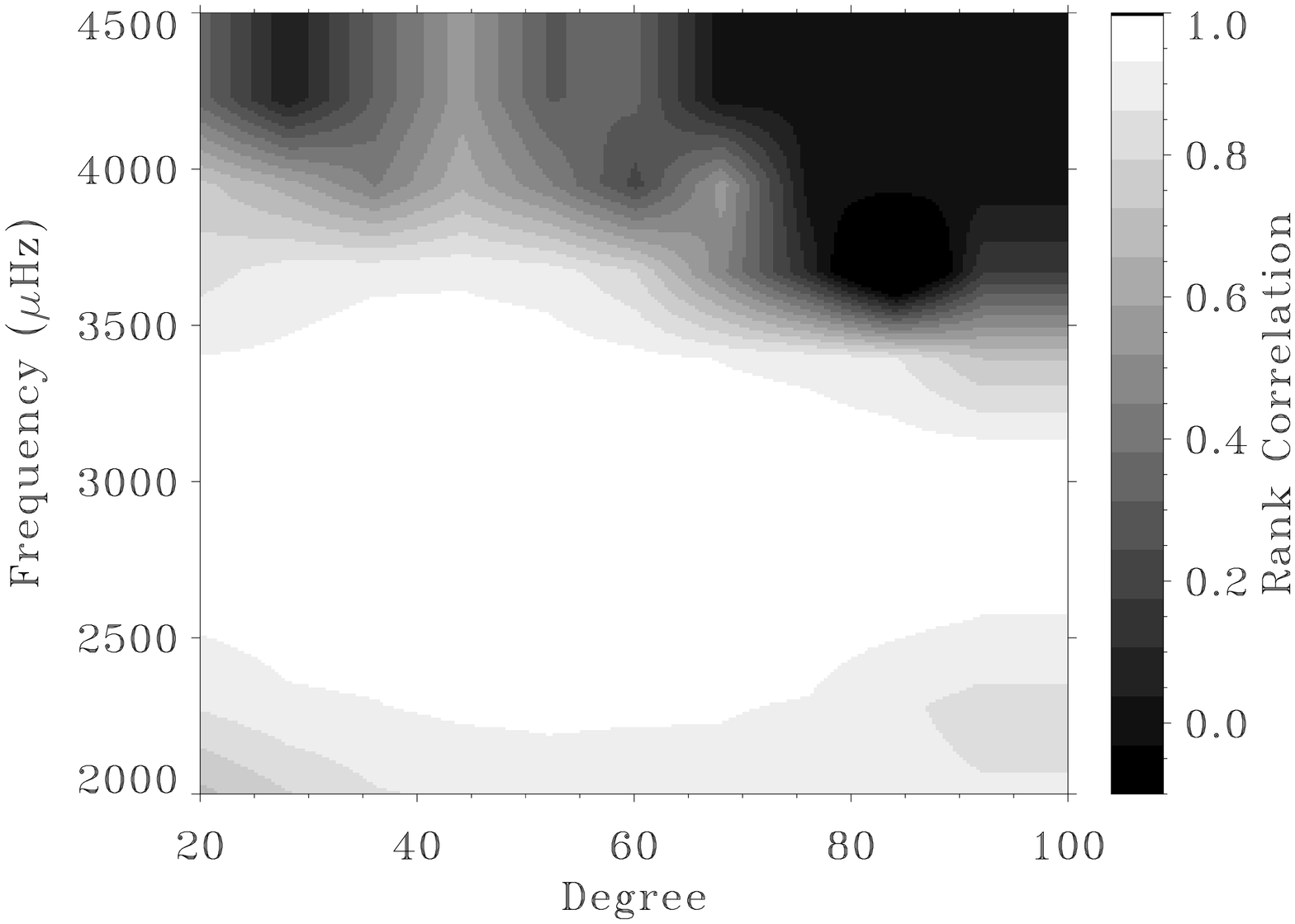,angle=90,width=12.cm,clip=}
\caption{Correlation maps between the frequency shifts of
individual modes and $\delta{\rm{MPSI}}$ 
as a function of  degree (\,$\ell$\,) and frequency (\,$\nu$\,).
The coefficients obtained from Spearman's rank
correlation analysis are binned 
with a bin size of eight in $\ell$ and $300~\mu$Hz 
in $\nu$. The top and bottom panels correspond to 
frequencies measured from nine days and 108 days duration, respectively. }
\label{cormap}
\end{center}
\end{figure}
\subsection{Correlation of Individual Modes} In order to investigate
the differences that exist between the low and intermediate-- degree
modes and between  different lengths of the observing run, we carry out
the same fitting and correlation analysis for each individual mode
using only the GONG data. The Spearman rank correlation coefficients
for $\delta{\rm{MPSI}}$ are shown in Figure~\ref{indicor}.  It is
evident that the correlation is a function of frequency and shows a
three-part structure: a rise in the low-frequency range, a plateau in
the five-minute band, and a decrease in the high-frequency zone.  In
the case of nine days (top panel of Figure~\ref{indicor}), the extent of
the plateau region is small indicating that a high value of correlation
exists only for a small mode set.  On  average, it appears that the
nine-day correlation coefficients are small  compared to the 108-day
coefficients confirming that the correlation is a function of the
length of the observation. We also note negative coefficients
indicating anti-correlation at low ($\nu <$ 1700 $\mu$Hz) and high
($\nu >$ 4200 $\mu$Hz) ends of the frequency range. Similar
anti-correlation has been reported by  Ronan, Cadora, and LaBonte
(1994), Harvey (1995), Jefferies (1998), and Rhodes, Reiter, and Schou
(2003).  However, the frequency range where the shift becomes negative
is different in our study than those reported earlier. One possible
explanation of the discrepancy could be the lack of sufficient modes at
higher frequencies in this analysis. Further work is needed to
understand the source of the difference and is beyond the scope of the
present paper.

To better illustrate the dependence of the correlation on mode
characteristics, we make rank correlation maps as a function of  $\ell$
and $\nu$ for all modes in nine-day and 108-day data sets with a bin
size of eight in $\ell$ and 300~$\mu$Hz in $\nu$
(Figure~\ref{cormap}).  In addition to the frequency dependence of the
correlation seen in Figure~\ref{indicor}, these maps show a decreasing
correlation with decreasing $\ell$ for the nine-day sets.  This
decrease is markedly smaller for the 108-day analysis. The effect may
arise from the increase of the mode lifetime with decreasing $\ell$.
Longer-lived modes will be observed less often in short time series due
to the greater temporal interval between excitation events. The
decrease of the signal-to-noise ratio at low frequencies and low
degrees may also play a role. Since the value of correlation decreases
at low-degree and low frequency, it seems possible that, for low-degree
modes ($\ell$ $\le$ 3), it may drop to the values observed by Chaplin
\etal\ (2001). Figure~\ref{cormap} emphasizes that the correlation
between frequency shifts and activity indices are complex and have
different behaviours for different mode sets and observing periods.

\section{Summary}
Using GONG time series over a period of ten years from 7 May 1995 to
16 October 2005, we have computed frequencies for periods of nine
days. The high quality data allowed us to study the frequencies of a
single mode and we demonstrate  that  individual multiplets show
significant temporal variations with the solar activity cycle. We find
that the frequency shifts exhibited by the nine-day time series are
consistent with the longer time series used for studies of internal
structure measurements and dynamics, and the frequency shift is
significantly correlated with different activity indices representing
different heights in the solar atmosphere.  Further, we show that the
correlation between frequency shift and the change in activity indices
varies on a yearly basis and is a function of the sensitivity factor as
measured from the slope of the linear fits. Although, the slopes do not
have a simple relation with the change in activity indicators, the
slopes and magnitude of the correlation between the mean frequency
shifts and mean activity measures are well correlated.  The analysis
further demonstrates that there are significant differences between
long-term and short-term variations and averages of short-term
variations may not accurately reflect the long-term variations and
vice-versa.

We also find clear evidence of the phase dependence of the shifts for
magnetic activity indices (MWSI and MPSI) showing significant
differences in behaviour of the magnetic flux emerging over the
ascending and descending phases and confirm the low-degree results of
Chaplin \etal{} (2001).  Although, there is no significant difference
in correlation between frequencies obtained from different observing
lengths, the sensitivity factor is higher for frequencies obtained from
short-duration time series agreeing with the result of Rhodes, Reiter,
and Schou (2003).  Finally, from the study of individual modes, we
infer that correlation between the shifts and activity indices is
complex.  More work is needed to understand and find a physical
mechanism that could explain the detailed behaviour of the frequency
shifts.

\acknowledgements
We thank the referee for comments which have improved the manuscript
substantially.  This work was supported by NASA grants NNG 5-11703 and
NNG 05HL41I.  This work utilizes data obtained by the Global
Oscillation Network Group (GONG) program, managed by the National Solar
Observatory, which is operated by AURA, Inc. under a cooperative
agreement with the National Science Foundation. The data were acquired
by instruments operated by the Big Bear Solar Observatory, High
Altitude Observatory, Learmonth Solar Observatory, Udaipur Solar
Observatory, Instituto de Astrof\'{\i}sica de Canarias, and Cerro
Tololo Interamerican Observatory.  This study also includes data from
SOHO/MDI.  SOHO is a mission of international cooperation between ESA
and NASA. This study includes data from the synoptic program at the
150-Foot Solar Tower of the Mt. Wilson Observatory.  The Mt. Wilson
150-Foot Solar Tower is operated by UCLA, with funding from NASA, ONR,
and NSF, under agreement with the Mt. Wilson Institute. We thank
Lawrence Puga for the Mg data.

\end{article} 
\end{document}